# Contagious obesity:
# from adenovirus 36 to RB dysfunction

**Razvan Tudor Radulescu**

Molecular Concepts Research (MCR), Muenster, Germany
E-mail: ratura@gmx.net

## ABSTRACT

Significant overweight represents a major health problem in industrialized countries. Besides its known metabolic origins, this condition may also have an infectious cause, as recently postulated. Here, it is surmised that the potentially causative adenovirus 36 contributes to such disorder by inactivating the retinoblastoma tumor suppressor protein (RB) in a manner reminiscent of a mechanism employed by both another pathogenic adenoviral agent and insulin. The present insight additionally suggests novel modes of interfering with obesity-associated pathology.

**Introduction**

Obesity has recently been qualified as a condition that is meanwhile posing a more significant threat to human health than undernutrition or infectious diseases (Kopelman 2000). Notably, such pathologic overweight is also a component of syndrome X, a group of diseases predicted to prevail in the 21st century (Reaven 2005).

As an etiological variant distinct from purely metabolic aberrations, a correlation between an adenovirus infection and human obesity was discovered about a decade ago (Dhurandar *et al.* 1997). This form of infectobesity (Pasarica & Dhurandar 2007), as it was also termed, has specifically been linked to human adenovirus 36 (Ad36) given that neutralizing antibodies to this agent were revealed in obese subjects and, moreover, Ad36 was shown to be capable of stimulating preadipocyte differentiation (Vangipuram *et al.* 2004). Interestingly, PCR data indicated that the Ad36 E1A gene was expressed in an Ad36-infected murine preadipocyte cell line serving as a model for exploring mechanisms of human obesity (Vangipuram *et al.* 2004). Subsequently, Dhurandhar and coworkers established that, in contrast to measurable Ad36 mRNA levels, the control non-adipogenic adenovirus 2 mRNA was not expressed, thus providing a putative clue as to why among these two closely related adenoviruses only Ad36 seems to possess adipogenic potential in humans (Rathod *et al.* 2007). In this context, it should be noted that while the E4 gene product appears to be important for certain adenoviral effects, as suggested by several experimental settings (O´Connor & Hearing 2000; Rogers *et al.* 2008), E4 expression is known to require the presence of E1A (Flint & Shenk 1989) which in turn likely represents an essential functional relationship during natural adenoviral infections.

**Conjecture and corollaries**

Based on my investigations on the potential role of the insulin-retinoblastoma protein (RB) complex in various human diseases including most recently obesity (Radulescu 2006; Radulescu 2007) and, moreover, given the fact that the insulin B-chain and Ad5 E1A protein share the LXCXE RB-binding motif (Radulescu &

Wendtner 1992), the present study addressed the question as to whether the Ad36 E1A protein contains an amino acid sequence related thereto which, if present, would significantly corroborate the involvement I had anticipated for RB (inactivation) in obesity.

Intriguingly, I have now identified an LXCXE motif spanning residues 105-109 of human Ad36 E1A protein, both in its 12S and 13S isoforms (Fig. 1).

**Leu** Thr **Cys** His **Glu**          Ad5 E1A$_{122\text{-}126}$

**Leu** Val **Cys** Gly **Glu**          human insulin B-chain$_{17\text{-}21}$

**Leu** Arg **Cys** Tyr **Glu**          human Ad36 E1A 12S/13S protein$_{105\text{-}109}$

Fig. 1   Alignment of LXCXE amino acid motifs- whereby X stands for any amino acid- in adenovirus 5 (Ad5) E1A protein, human insulin B-chain and human adenovirus 36 (Ad36) E1A protein.

This finding suggests that, similar to insulin (Radulescu & Wendtner 1992; Radulescu *et al.* 1995; Radulescu *et al.* 2000; Radulescu & Schulze 2002; Radulescu 2006; Radulescu & Kehe, 2007; Radulescu 2007), the Ad36 E1A protein may bind and thereby inactivate RB which in turn could contribute to the genesis of infectobesity.

If this conjecture was experimentally validated, e.g. by demonstrating significant amounts of Ad36 E1A-RB complexes in (pre)adipocytes *vs.* control cells, then it may be worthwhile exploring as to whether blocking such Ad36 E1A-RB complex formation, e.g. by means of LXCXE motif-"chelating" MCR peptides (Radulescu *et al.* 2000; Radulescu & Kehe, 2007), could attenuate or even reverse this metabolic disorder of presumed infectious origin.

Conversely, the Ad36 E1A-insulin similarity unveiled here lends further support to the notion according to which the insulin-RB complex is likely to be crucially involved not only in the pathophysiology of cancer, but also of (both non-viral and viral forms of) obesity (Radulescu 2006; Radulescu 2007).

Consistent with this proposed dysfunction of the key inhibitor of cell cycle progression RB during pathologic adipogenesis, it has recently been shown that abrogating fatty acid synthase function induces cell cycle arrest (Knowles *et al.* 2004).

Taken together, inhibition of the growth-suppressive activity of RB through its direct binding to functionally related factors such as insulin and/or Ad36 E1A, as I am suggesting here, could turn out to decisively contribute to an "obesogenic" environment that is known to predispose to the development of a variety of human malignancies, but has so far been poorly understood (Renehan *et al.* 2008). If so, viruses would have again enlightened us.